\documentstyle[prd,aps,preprint,eqsecnum]{revtex}
\begin{document}

\title {Quark-like potentials in an extended Maxwell theory}

\author{Harry Schiff\thanks{E-mail:~hschiff@phys.ualberta.ca}}

\address{ Department of Physics, Theoretical Physics Institute,
University of Alberta\\
Edmonton, Alberta,   T6G 2J1.}

\maketitle

\begin{abstract}
The exact Li$\acute{e}$nard-Wiechert solutions for the point charge in 
arbitrary motion are shown to be null fields everywhere. These are used
as a basis to introduce extended electromagnetic field equations that have null 
field solutions with fractional charges that combine with absolute confining 
potentials.
\end{abstract}

\newpage

\section{Introduction}

Some time ago I pointed out~\cite{H} that the Li$\acute{e}$nard-Wiechert
solutions in the Lorentz gauge for a point charge $g$ in arbitrary
motion, the following relation between fields and potentials holds
everywhere
\begin{equation}
  F^{2}_{\mu \nu} = -2g^{-2}(A^{2}_{\mu})^{2} ,
\end{equation}
where~\cite{R} $F_{\mu \nu} \equiv \partial_{\mu}A_{\nu} -
\partial_{\nu}A_{\mu},\; x_{4}$ = it, so that
$$
  F^{2}_{\mu \nu} = 2(H^{2} - E^{2}),\; A^{2}_{\mu} = A^{2} - \phi^{2}.
$$
By looking at (1.1) in the rest frame of $g$ it is clear that
it applies only to coulomb solutions.
In addition, the exact solutions satisfy $\bf{E \cdot H} = 0$, i.e.,
\begin{equation}
F_{\mu \nu}F_{\mu \nu}^{*} = 0.
\end{equation}
$$
{\rm Writing}\;(1.1)\;{\rm as}\;\;\;\;\;\;\;\;\;\;\;\;\;\;\;\;\;\;\;\;\;\;
\qquad F^{2}_{\mu \nu} + 2g^{-2}(A^{2}_{\mu})^{2} = 0 , 
\;\;\;\;\;\;\;\;\;\;\;\;\;\;\;\;\;\;\;\;\;\;\;\;\;\;\;\qquad\qquad\qquad (1.1)`
$$
shows that it can be expressed as the square,
\begin{equation}
  [F_{\mu \nu} + \sqrt{2}g^{-1}A_{\mu}A_{\nu}]^{2} = 0
\end{equation}
due to the opposite symmetries of $F_{\mu \nu}$ and
$A_{\mu}A_{\nu}$. Defining the mixed tensor
\begin{equation}
G_{\mu \nu} \equiv  F_{\mu \nu} + \sqrt{2}g^{-1}
A_{\mu}A_{\nu}
\end{equation}
and noting that $F_{\mu \nu}^{*} = G_{\mu \nu}^{*}$, $(1.2)$ and  $(1.1)`$
can be written as 
\begin{equation}
G_{\mu \nu}G_{\mu \nu}^{*} = 0,
\end{equation}
\begin{equation}
G_{\mu \nu}G_{\mu \nu} = 0.
\end{equation}
Consequently $G_{\mu \nu}$ satisfies the conditions for a null field, 
which hold everywhere for the exact solutions, compared to more general 
solutions of Maxwell's equations for which only the radiation field is
null. \\
\indent The null field $G_{\mu \nu}$ is interesting, stemming as it does from 
an exact solution to Maxwell's equation, but is that its only significance 
or is there more, perhaps, that can be explored? This is the motivation for
considering the non-gauge invariant equation,
\begin{equation}
  \partial_{\mu}G_{\mu \nu} = 4 \pi j_{\nu} \qquad i.e.,
\end{equation}
\begin{equation}
  \partial_{\mu}(F_{\mu \nu} + \sqrt{2}g^{-1} A_{\mu}A_{\nu}) = 4\pi j_{\nu},
\end{equation}
where $j_{\nu}$ is a localized 4-current. The total current thus consists
of $j_{\nu}$ plus $-\sqrt{2}g^{-1}\partial_{\mu}A_{\mu}A_{\nu}$, an intrinsic 
contribution from the potentials, which are now physically significant since gauge 
invariance does not apply.\\
\indent In this note I examine the simplest solution of (1.8), static and radially 
symmetric, which also satisfy the null conditions (1.5) and (1.6).  

\section{Solutions of (1.8)}

For all solutions considered here, $A = A_{r}(r)$, thus ${\bf H} = 0$ and the null
condition (1.5) is identically satisfied. The magnetic part of (1.8) is then,
\begin{equation}
\sqrt{2}g^{-1}\nabla {\bf \cdot} ({\bf A}A) = 4 \pi J(r),
\end{equation}
with the simple solution of the homogeneous part,  
\begin{equation}
  A = k/r,
\end{equation}
\noindent where $k$ is arbitrary.\\
The linear electric field equation from (1.8) can be written 
\begin{equation}
  \nabla \cdot {\bf E}\; - \;\sqrt{2}g^{-1}  \nabla  \cdot ({\bf A}
  \phi) = 4 \pi \rho(r).
\end{equation}
\noindent The homogeneous part of (2.3) becomes, with
\begin{equation}
                           \gamma \equiv \sqrt2 g^{-1}k,
\end{equation}
\begin{equation}
                                A = \gamma g/\sqrt{2}r,
\end{equation}
\begin{equation}
  \frac{1}{r} \frac{d^{2}(r\phi )}{dr^{2}} +
  \frac{\gamma}{r^{2}} \frac{d(r\phi )}{dr} = 0 .
\end{equation}
For any value of $\gamma$, except $\gamma = 1$, there are two
indicial solutions of (2.6),
\begin{equation}
\phi_{1} = {b \over r}
\end{equation}
\begin{equation}
\;\;\;\;\;\;\;\phi_{2} = c_{2}r^{-\gamma},
\end{equation}
\noindent consisting of a Coulomb potential and, for $c_{2} > 0,\; \gamma < 0,$
an absolute confining potential.\\
\indent The total charge of the source in (2.3),  using (2.7) and (2.8) with an application
of the divergence theorem, is 
\begin{equation}
\int \rho d^{3}x = (1 - \gamma)b
\end{equation}
\noindent and is independent of the confining amplitude $c_{2}.$\\
Also, for any value of $\gamma$ the total source current in (2.1), using the divergence 
theorem, is 
\begin{equation}
\int J(r)d^{3}x = \gamma^{2}g/\sqrt2.
\end{equation}   
\indent For $\gamma = 1$ the two indicial solutions of (2.6) merge to a single Coulomb solution,
so a second solution is needed. This is easily seen to be given by $r\phi \sim logr.$  The
two independent solutions for $\gamma = 1$ are thus, 
\begin{equation}
\phi_{3} = {b \over r}
\end{equation}
\begin{equation}
\phi_{4} = \frac{c_{4} log (r/\alpha)}{r}.
\end{equation}   
The logarithmic solution (2.12) can represent two possible confining potentials
for $c_{4} > 0$ and $c_{4} < 0,$  but not absolute ones. For $c_{4} > 0, \; \phi_{4} \rightarrow - \infty$
as $r \rightarrow 0$ and peaks at $r = \rm{e}\alpha,$  while for $c_{4} < 0,\; \phi_{4} \rightarrow  \infty$
as  $r \rightarrow 0$, then forms a well of finite depth. For both cases $\phi_{4} \rightarrow 0$ as 
 $r \rightarrow \infty.$  Although the Coulomb potential (2.11) is shown separately it can be 
absorbed in (2.12) by a redefinition of $\alpha$.\\
\indent The total charge of the source (2.3) for $\gamma = 1$, using (2.11) and (2.12), with the application of 
the divergence theorem, yields 
\begin{equation}
\int \rho\ d^{3}x = - c_{4}
\end{equation}   
\noindent and is independent of the Coulomb potential. However, due to the fact that (2.12) is not 
absolutely confining as well as the unsatisfactory charges resulting from $\gamma = 1$, indicated
below, (2.12) will not be considered further.\\
\indent Below, the null condition (1.6), or equivalently (1.1)`, is applied using (2.5) and (2.7) for
$\gamma < 0$, which is possible because both $A$ and $\phi$  go as $1/r$.

\section{Null Fields for ${\bf\gamma} < 0$}

For the static, radially symmetric solutions of (1.8) the null condition (1.1)` becomes
\begin{equation}
E^{2} + g^{-2}(A^{2}-\phi^{2})^{2} = 0,
\end{equation}
\noindent leading to the quadratic equation for the Coulomb charge $b$ in (2.7),
\begin{equation}
b^{2} \pm gb - \gamma^{2} g^{2}/2 = 0.
\end{equation}
\noindent To avoid one set of $\pm$ signs from (3.2), the minus sign is chosen for solutions
shown below; since $b$ changes sign with $g$ the remaining solutions are obtained directly
for the $+$ sign in (3.2). Thus, 
\begin{equation}
b^{2}_{-} -  gb_{-} - \gamma^{2}\! g^{2}/2 = 0
\end{equation}
with solutions
\begin{equation}
 b_{-}(1,2) = g/2[1 \pm \sqrt{(1 + 2\gamma^{2})}].
\end{equation}
For the two solutions $b_{-}(1)$ and $b_{-}(2)$ of (3.4) the following relations are seen to 
apply,
$$b\quad \propto \quad g$$
\begin{equation}
b_{-}(1) +  b_{-}(2) = g,
\end{equation} 
\begin{equation}b_{-}(1) -  b_{-}(2) = g \sqrt{(1 + 2\gamma^{2})},
\end{equation}
\begin{equation}
b_{-}(1)/b_{-}(2) = [1 + \sqrt{(1 + 2\gamma^{2})}]/[1 -  \sqrt{(1 + 2\gamma^{2})}].
\end{equation}
\noindent Choosing $b_{-}(1)/b_{-}(2) = -2$ and $b_{-}(1) = 2/3$ one gets $\gamma^{2} = 4, \;
g = 1/3$ and obviously $ b_{-}(2) = - 1/3$. Hence all solutions of (3.2) are,
\begin{equation}
b_{-}(1,2) = \pm(2/3, -1/3),\qquad  b_{+} = \mp(2/3, -1/3) 
\end{equation}
\noindent the results of the null fields solutions of (1.8) with $g = \pm 1/3$ and $\gamma = -2,$
corresponding to a quadratic confining potential (2.8).\\
\indent Null fields solutions also exist for which the charges $\pm 2/3, \;\pm 1/3$ appear
singly, paired with charges of varying values. Below is a list of some of these results 
for corresponding values of g and $\gamma $.

\begin{eqnarray}
g & = & \pm 1/3,\;\;\;\;\gamma  =  -2\sqrt{3},\;\;\;\;\;\;\;\;\;    b  =  (\mp/\pm)(1, -2/3) \nonumber \\ 
g & = & \pm 2/3,\;\;\;\;\gamma  =  -\sqrt{3/2},\;\;\;\;\;\;\;    b  =  (\mp/\pm)(1,-1/3)   \nonumber \\
g & = & \pm 2/3,\;\;\;\;\gamma  =  -2  \qquad \;     \;\;\;\;\;\;\; b =  (\mp/\pm)(4/3,-2/3)  \nonumber \\ 
g & = & \pm 1,\;\;\;\;\;\;\;\; \gamma  =  -2\sqrt{2/3},\;\;\;\;\;  b  =  (\mp/\pm)(4/3, -1/3) 
\end{eqnarray}
Additional null fields exist for charges $\pm2/3, \ \pm1/3$ with paired 
charges greater than $4/3$. ( With regard to $\gamma = 1$ the ratio of charges from (3.7) has
the unsatisfactory values of $\pm(1 + \sqrt3)/(1 - \sqrt3)$.)

\section{Another `Solution'}

 In this section, another solution asymptotic in nature, is considered. 
Referring to (2.3), with $\rho = 0$, a solution evidently exists with $\phi$
a constant and $\nabla \cdot {\bf A} = 0$. With the possibility that this is 
an asymptotic form to a more general solution for which  $\nabla \cdot
 {\bf A} = 0$,
a solution to (2.3) is sought with
\begin{equation}
A = \beta/r^{2},
\end{equation}
where $\beta$ has dimension of length times charge.

Although (4.1) is not a solution of the homogeneous part of (2.1), it will approach it 
asymptotically. Inserting it into the left side of (2.1) gives
\begin{equation}
\sqrt{2} g^{-1}  \nabla \cdot ({\bf A}A) = \sqrt{2} g^{-1} \beta^{2}[-2/r^{5} +
4\pi \delta ({\bf r}) / r^{2}].
\end{equation}
\noindent With (4.1) the electric field equation (2.3) becomes 
\begin{equation}
  \frac{d}{dr} (\frac{d\phi}{dr}) + \frac{2}{r} \frac{d\phi}{dr} =
  -\frac{\sqrt{2} g^{-1} \beta}{r^{2}} \frac{d\phi}{dr} - 4\pi \delta(0)
  \phi(0) - 4\pi \rho(r).
\end{equation}
\noindent Dividing the homogeneous part by $d\phi/dr$ and integrating, one
finds 
\begin{equation}
  -\frac{d\phi}{dr} = E = \frac{q}{r^{2}}\exp\left({\sqrt{2} g^{-1} \beta \over r}\right),
\end{equation}
\noindent where $q$ is a constant of integration giving the potential
\begin{equation}
  \phi = \frac{gq}{\sqrt2\beta}\exp\left({\sqrt{2} g^{-1} \beta \over r}\right),
\end{equation}
\noindent making the obvious choice $\sqrt{2} g^{-1}\beta < 0$.

Asymptotically  
\begin{equation}
  \phi \simeq \frac{gq}{\sqrt2\beta} + \frac{q}{r} +
  \frac{\sqrt2 g^{-1} q\beta}{r^{2}} + \quad  \ldots ,
\end{equation}
exhibiting a Coulomb potential $q/r$, plus the constant $gq/\sqrt2\beta$.

Since $\phi(0) = 0$ in (4.5), the corresponding expression in the inhomogeneous term
of (4.3) is $0$. Also, applying the divergence theorem to (2.3) with the asymptotic
expression (4.6), shows consistency with the charge density $\rho$ being $0$.

The third term in (4.6), (which stretches the approximation) suggests a possible 
interpretation of a charge distribution, representing  perhaps a penetration at the
edge of a particle.

It is interesting that a charge from the 'quark' region, $g$, is reflected in a constant 
that appears in the asymptotic region; for a particle anti particle pair the two associated 
constants would naturally cancel. One could raise the question, should the asymptotic constant 
appear in some particle reaction, whether it could influence the results of the reaction.

\section{Conclusions}

The extended electromagnetic equations (1.8) have the important property of having Coulomb 
solutions combined with absolute confining potentials. In addition, the Coulomb fields in the 
subset of null field solutions have charges with quark-like fractional values $\pm2/3$ and 
$\pm 1/3$, which occur with specific values of the index in the confining potential. Thus the
non-gauge invariant mixed tensor $G_{\mu\nu}$, through the extended Maxwell equations (1.8), is 
seen to be more significant than being only a null field of the Li$\acute{e}$nard-Wiechert 
solutions. The value of unity for the index has been  used phenomenologically; the results 
obtained here analytically though,
 have definite values of the index. It would be of interest to see whether 
these values are signinficant.

A feature of the approximate asymptotic solution (4.6) is the appearance of a constant in the asymptotic 
region, which involves the basic charge $g$ in (1.8). A constant potential is irrelevant in the
Maxwell case, but in a transition between the `quark' region and the Maxwell domain, as seems to be 
suggested by the asymptotic solution, the effect of a constant may be relevant in some reactions.

It is an open question at present why the null fields are the important ones in the solutions of (1.8)
 for obtaining the familiar fractional charges, a more detailed treatment involving the dynamics
of the system and its quantization may shed some light on this question.    

\section*{Acknowledgments}
I want to thank J. David Jackson for helpful comments and suggestions,
also Werner Israel  for useful discussions.

\end{document}